\documentclass[sigplan,10pt]{acmart}

\startPage{1}

\setcopyright{none}

\usepackage{booktabs}   
\usepackage{subcaption} 
\usepackage{syntax}

\usepackage{tikz}
\usetikzlibrary{arrows,positioning} 
\tikzset{
    >=stealth',
    punkt/.style={
           rectangle,
           rounded corners,
           draw=black, very thick,
           text width=6.5em,
           minimum height=2em,
           text centered},
    pil/.style={
           ->,
           thick,
           shorten <=2pt,
           shorten >=2pt,}
}

\usepackage{xcolor}
\usepackage{algorithm}
\usepackage[noend]{algpseudocode}

\definecolor{uwpurple}{RGB}{128,0,128}

\begin{document}

\title[Qualifying Evaluation Report]{Using human-in-the-loop synthesis to author functional reactive programs}         


\author{Julie L. Newcomb}
\orcid{nnnn-nnnn-nnnn-nnnn}             
\affiliation{
  \department{Paul G. Allen School of Computer Science and Engineering}              
  \institution{University of Washington}            
  \city{Seattle}
  \state{WA}
}
\email{newcombj@cs.washington.edu}          

\author{Rastislav Bodik}
\orcid{nnnn-nnnn-nnnn-nnnn}             
\affiliation{
  \department{Paul G. Allen School of Computer Science and Engineering}              
  \institution{University of Washington}            
  \city{Seattle}
  \state{WA}
}
\email{bodik@cs.washington.edu}          

\begin{abstract}
Programs that respond to asynchronous events are challenging to write; they are difficult to reason about and tricky to test and debug. Because these programs can have a huge space of possible input timings and interleaving, the programmer may easily miss corner cases. We propose applying synthesis to aid programmers in creating programs more easily and with a higher degree of confidence in their correctness. 

We have written an efficient encoding of functional reactive programming (FRP) semantics based on functional programming over lists lifted in Rosette. We demonstrate that this technique is state-of-the-art by first comparing its performance against two existing synthesis tools that produce list manipulation programs, and then by synthesizing a suite of benchmarks given complete specifications. We also propose an interactive tool in which a programmer provides some initial partial specification in the form of input/output examples or invariants; the tool finds ambiguity in the specification by synthesizing two candidate programs and gives the user an input that distinguishes them; the user updates the specification and continues iterating until the correct program is found. As evaluation, we demonstrate the use of the tool on a suite of benchmarks from the web programming and Internet of Things domains and walk through a sample interaction on a realistic web benchmark, showing that we can converge on the target program with a tractable number of interactions. 

As future work, we discuss encoding additional FRP languages to in order to explore metalinguistic features, strategies for decomposition that would allow the synthesis of larger programs, and improved programmer tools such as a GUI to more easily elicit specifications.
\end{abstract}




\maketitle

\section{Introduction}

Programs that respond to asynchronous events, such as web applications and Internet of Things programs, pose a particular challenge to programmers. When writing an ordinary function, the programmer must consider all possible inputs---a program with integer inputs may need to consider how to handle positive, negative, and zero-valued inputs, or a program that takes lists as input may need to write a special case for the input of an empty list. Programs that operate over asynchronous events must not only anticipate any possible values for the events they receive, but also any possible timing or interleaving of events that might occur. The space of possible events considered over time is thus very large and it is quite likely that corner cases the programmer has not considered will occur. Furthermore, asynchronous programs are tricky to debug, especially without specialized tool support. Even with testing tools like Selenium, reproducing bugs in web applications is often challenging. Internet of Things programs, which have a far less developed tool ecosystem, may require users to physically walk past motion sensors, wait for a particular time of day, or otherwise induce real-world interactions in order to test their scripts.

For example, consider this faulty Internet of Things program: an IoT hub maintains a mode of `home' when the user is at home (determined by a geofenced mobile device), `away' when they are not at home, and `night' between dusk and dawn (determined by daily sunset and sunrise times). The user wants to turn on a kitchen light when a motion sensor goes off; further, they want the color of the light to be white during the day and orange at night. After setting up this program, the user leaves the house in the morning, changing the mode to `away'); at 7pm the sun sets and the mode shifts to `night'; the user returns at 8pm, setting the mode to `home'; finally, the user enters their kitchen at 11pm, and the kitchen light turns on in a white color. The bug lies in the choice of mode when the user arrives home during nighttime hours; most likely when the program was written, the user assumed they would return home before the sun set and failed to consider the case in which those events occurred in the other order.

To address this problem, we propose a synthesis tool that collaboratively defines a reactive program with a human programmer through an interactive loop. Rather than writing programs, the programmer will write specifications; however, specifications are always challenging to write. Furthermore, programmers often do not fully understand the specification of their program when they sit down to write code; they often flesh out their design or discover corner cases as they work. Thus, our tool asks the user only for an initial, partial specification. These may be in the form of input/output traces (programming by demonstration) or invariants that should hold over program inputs, outputs, or both. The tool then synthesizes two programs that satisfy the specification and asks the user to provide additional specifications to disambiguate them. This process repeats until only one semantically unique program can be found, or until the user is satisfied with the result. Because we synthesize programs over symbolic inputs that represent all possible event values and event timing within a given bound, we can offer a higher level of confidence that any corner cases have been fleshed out and that the final program is correct. Here we consider web programming and Internet of Things programs as our target domains, but there are many other domains that would be a good fit for this approach, such as game programming or robotics.

Our main contributions are: an efficient encoding of an FRP language and interpreter in Rosette; an encoding of real-world events as discretized lists of values; and an interactive synthesis tool that allows users to iteratively refine their specifications until the tool converges on the desired synthesized program.

\section{The reactive programming model}

There are several existing programming paradigms for programs that respond to asynchronous events. Javascript's event handlers are notoriously difficult for programmers to work with (often referred to as ``callback hell''). Internet of Things programs are most often written in sets of event-condition-action rules, but their semantics are not well defined~\cite{newcomb2017iota} and no clear standard for programming these devices has emerged. Instead, we chose functional reactive programming (FRP) as our target synthesis language. Functional reactive programming is an approach to reactive programming that applies functional programming techniques to elements whose values can vary over time. One widely-used FRP framework is Microsoft's Reactive Extensions (Rx), which has been implemented in many languages including C++, Scala, and Python. Recently FRP has been gaining in popularity for web programming, with the Elm language\cite{czaplicki2012elm} and Javascript libraries such as cycle.js and bacon.js. FRP provides a clean, well-defined abstraction that can easily be lifted to Rosette.  This work takes the Javascript-based Flapjax language~\cite{meyerovich2009flapjax} as its starting point. Flapjax is a fairly ``classical'' FRP language with support for both event streams (discrete events that occur over time) and behaviors (continuous values that can vary over time). Flapjax's relatively small API and its composable syntax also make it well suited as a target language for synthesis.

\begin{figure}
    \centering
    \begin{tikzpicture}[node distance=1cm, auto,]

 \node[punkt] (mouse-down) {Mouse down events};
 \node[punkt, inner sep=5pt,below=0.5cm of mouse-down]
 (constantE-t) {constantE \#t}  edge[pil,<-] (mouse-down);
 
 \node[punkt, right=1cm of mouse-down] (mouse-up) {Mouse up events};
 \node[punkt, inner sep=5pt,below=0.5cm of mouse-up] 
 (constantE-f) {constantE \#f} 
 edge[pil,<-] (mouse-up);
 
 \node[right=0.5cm of mouse-down] (dummy) {};
 
 \node[punkt, below=2cm of dummy] (mergeE) {mergeE}
 edge[pil,<-,bend left] (constantE-t)
 edge[pil,<-,bend right] (constantE-f);
 
 \node[punkt, below=0.5cm of mergeE] (startsWith) {startsWith \#f}
 edge[pil,<-] (mergeE);
 
 \node[punkt, below=2.75cm of constantE-f] (mouse-move) {Mouse move events};
 
 \node[punkt, below=4cm of constantE-t] (snapshotE) {snapshotE}
 edge[pil,<-, bend left] (startsWith)
 edge[pil,<-, bend left] (mouse-move);
 
 \node[punkt, below=4cm of mergeE] (ifE) {ifE}
 edge[pil,<-,bend right=55] (snapshotE)
 edge[pil,<-,bend left=50] (mouse-move)
 edge[pil,<-,bend left=45] (snapshotE);
 
 \node[punkt, below=0.5 of ifE] (filterE) {filterE identity}
 edge[pil,<-] (ifE);

\end{tikzpicture}
    \caption{The dataflow graph underlying the drag and drop program.}
    \label{fig:ddgraph}
\end{figure}
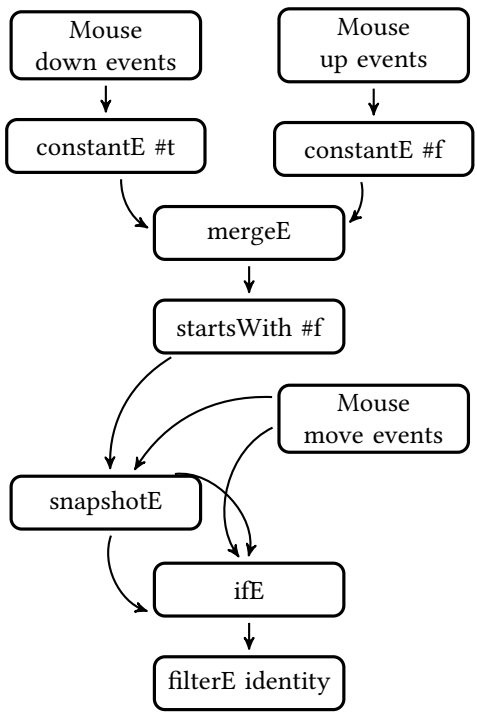

As a running example we will use a simple drag and drop program from the web programming domain. A user clicks on the target element, moves their mouse to move the element, and releases the mouse button to drop the element in the new desired position. Consider the Flapjax implementation of the program in figure~\ref{fig:ddfj} and the dataflow graph underlying that program in Figure~\ref{fig:ddgraph}. The program takes three event streams from the HTML document: a stream of mouse-down events that occur when the user holds the mouse button down on the target element, a stream of updated mouse position coordinates events that occur every time the user moves the mouse, and a stream of mouse-up events that occur whenever the user releases the mouse button. The output will be a stream of updated coordinates for the dragged element; code to update the DOM with the element's new position is omitted for clarity

First, we create a behavior \texttt{is\-draggingB} that will be the continuous boolean value indicating whether the element is currently being dragged or not. This is done by replacing the mouse-down events with the boolean value \texttt{true} and the mouse-up event with the boolean value \texttt{false}. These two event streams are merged into one with the \texttt{mergeE} operator and then cast into a behavior with \texttt{startsWith} (since behaviors must always have a value, this operator gives it the initial value of \texttt{false}). Now, every time a new mouse coordinate event occurs, we want to check if we are currently in a dragging state or not. We use the \texttt{snapshotE} operator to sample the value of the dragging state behavior \texttt{is\-draggingB} when a new mouse-move event is emitted. Finally, we create an event stream that contains only mouse-mouse coordinates while the program is in a dragging state with \texttt{ifE} by using the sampled \texttt{is\-draggingE} values as a guard, returning the mouse-move coordinates if the dragging state is true and false otherwise. We add a \texttt{filterE} operator to return the false-valued events and now have an event stream containing coordinates that should be used to change the position of the target element.

\begin{figure*}
\centering
\begin{verbatim}
function updateEltCoordinates() {
 var downE = extractEvent(target,"mousedown");    // stream of mouse down events
 var upE = extractEvent(document,"mouseup");      // stream of mouse up events
 var moveE = extractEvent(document,"mousemove");  // stream of mouse position coordinates

 var is-draggingB = downE.constantE(false)       // a behavior that is true when the mouse button is
                    .mergeE(upE.constantE(true)) // held down and false otherwise
                    .startsWith(false);                                          
 var is-draggingE = moveE.snapshotE(is-draggingB); // a stream of dragging state status at the time of
                                                   // each update of mouse position
                                                   // a stream of mouse position if in dragging state
 return is-draggingE.ifE(moveE,is-draggingE)       // and false otherwise
                .filterE(function(x) { return x; });   // filter to remove false events
}                                                      // returns stream of new element positions
\end{verbatim}
    \caption{Drag and drop written in Flapjax (Javascript). The code to update the position of the target element with the function return value is omitted.}
    \label{fig:ddfj}
\end{figure*}

\section{Encoding an FRP interpreter}

To synthesize these programs, we use Rosette\cite{torlak2014lightweight}, an extension of Racket with a symbolic virtual machine that translates lifted Racket programs into constraints that can be passed to a solver. Thus, we need to define an encoding that can represent the semantics of our FRP language in Rosette's lifted subset of Racket. We build up our encoding in stages: first, we discuss using Rosette to model functional programming over lists; next, we discuss our encoding for event streams and behaviors; finally, we detail how program sketches can be defined for our DSL.

\subsection{Functional programs over lists}

As a first step, we consider non-reactive functional programming over the list domain. We consider the DSL used in the Deepcoder tool\cite{BalGauBroetal16}, which synthesizes programs using a variety of search strategies enhanced by machine learning (see the sample program in Figure~\ref{fig:sampledcdslprogram}; the full grammar for the DSL is in Appendix~\ref{sec.dcdslgrmr}). This DSL is a set of functional combinators over integers and lists of integers; some of these combinators also take predicate functions as arguments. With our encoding of the DSL, we can execute programs against concrete inputs. This same encoding can be applied to symbolic inputs with no additional effort. Combinators in our DSL can take integers, lists of integers, and predicate functions as inputs. We represent integer inputs as Rosette symbolic integers and list inputs as lists of symbolic integers with fixed length. 

The semantics of each combinator is represented through a Racket function, largely taking advantage of the list methods lifted by Rosette. As an example, consider the \texttt{scanl1} operator, with Rosette implementation given in Figure~\ref{fig:scanl1}. This operator acts as a fold over each prefix of the input list. Our encoding returns an empty list if the input list is empty, and otherwise uses a for loop to calculate the fold of each prefix of the list, returning a list of the same length as the input. The operator can be applied to concrete inputs as well as symbolic ones; when the operator is applied to a list of symbolic integers \texttt{(list \$i0 \$i1 \$i2)} and the prediction function \texttt{+}, Rosette's symbolic evaluation returns the following formula:

\begin{verbatim}
(list i$0 (+ i$0 i$1) (+ i$2 (+ i$0 i$1)))
\end{verbatim}

For higher-order combinators such as \texttt{scanl1}, \texttt{map}, and \texttt{filter}, we need to define predicate functions as well. The DSL includes a small set of predicate functions with signatures integer->integer, integer->boolean, and integer->integer->integer. Most of these functions can be straightforwardly modeled as well, but some, such as multiplication, division by constant, and remainder operators, are undecidable under the theory of integers and thus cannot be solved by Rosette efficiently. We handle these functions by defining lookup tables over bounded inputs, taking advantage of arithmetic properties to reduce the size of these tables wherever we can. For example, we only store the squared values of positive inputs, since they will always be the same as their negative counterparts; we also sort inputs before calling multiplication to exploit the fact that the operation is commutative. Of course, these lookup table functions are only semantically equivalent to the functions they encode over bounded inputs, so care must be taken to stay within those bounds when synthesizing programs.

\begin{figure}
    \centering
    \begin{verbatim}
(define (scanl1 f xs)
  (if (empty? xs)
      '()
      (for/list ([i (range (length xs))])
        (foldl f (first xs) (take (drop xs 1) i)))))
    \end{verbatim}
    \caption{The \texttt{scanl1} operator implemented in our Rosette encoding of the list manipulation DSL.}
    \label{fig:scanl1}
\end{figure}

\begin{figure}
    \centering
    \begin{verbatim}
(define (program int int-list)
  (define r1 int)
  (define r2 int-list)
  (define r3 (sort r2))
  (define r4 (take r1 r3))
  (define r5 (sum r4))
  r5)
    \end{verbatim}
    \caption{An example program in the Deepcoder DSL.}
    \label{fig:sampledcdslprogram}
\end{figure}

Given an encoding of the DSL, we are able to perform verification; for example, we can test that two programs are equivalent over inputs of some bounded size. In order to do synthesis, we will need to implement sketches in order to define the program search space.

\subsection{Sketches}

With the semantics of the DSL fully modeled, we can produce sketches that represent programs within the DSL. For reasons of solver efficiency, we represent programs as a series of register definitions. Inputs to the program sketch are assigned to registers; for every subsequent instruction, we take one or more previously defined registers, apply them to a combinator, and store the resulting value in a new register. Each instruction can thus be represented by a set of indexes into lists: an index into the list of operators to indicate which operator to apply, indexes into the lists of previously computed registers, and optionally indexes into the list of predicate functions. As the Deepcoder DSL combinators can return either integers or lists of integers, we store previously computed registers in two typed arrays. The value of the last register computed becomes the return value of the program. The number of instructions in each sketch is fixed, but since the program is not constrained to use all computed registers to calculate the final return value, instructions that calculate values that are not used later in the program are essentially no-ops and the fixed number of instructions in a sketch is essentially an upper bound. Since each instruction is represented by a set of symbolic integers which act as indexes into concrete lists, they can be synthesized very efficiently since the possible values is very constrained. 

\begin{figure}

\begin{tabular}{cc}
\begin{tabular}{|c|c|}
\hline
    index & operator name \\
\hline
    0 & filterE \\
    1 & constantE \\
    2 & delayE \\
\hline
\end{tabular}

&

\begin{tabular}{|c|c|}
\hline
  index  & predicate function \\
\hline
   0 & ($\lambda$ (i) (= i 3)) \\
   1 & ($\lambda$ (i) (> i 3)) \\
   2 & ($\lambda$ (i) (< i 3)) \\
\hline
\end{tabular}
\end{tabular}
\vspace{5mm}

\begin{tabular}{|c|c|c|c|c|}
\hline
 Register & Op code & Input & Function & Integer \\
\hline
 R1 & 2 & 0 & - & 3 \\
 R2 & 0 & 1 & 0 & - \\
 R3 & 1 & 2 & - & 2 \\
\hline
\end{tabular}
\caption{A simplified example of a register-style concrete program sketch. The sketch encodes the program \texttt{input.delayE(3).filterE(function(i) \{ return i == 3; \}).constantE(2);}. The op codes are indexes into the table of operators; the Function value in R2 is an index into the predicate function table.}
\label{fig:toysketch}
\end{figure}

\subsection{Encoding of event streams and behaviors}

With a satisfactory encoding for synthesizing functional programs over list inputs, we next turn to the problem of encoding asynchronous inputs. We need to be able to represent event streams (series of discrete events) and behaviors (continuous values that vary over time) over a bounded period of time. Lists are a very natural data structure for this purpose. We considered two means of encoding our two datatypes; examples in both encodings are given in Figure~\ref{fig:encodings}. First, we considered representing event streams as pairs of timestamps and values, and behaviors as an initial timestamp-value pair, followed by pairs of timestamps and values for every point at which the behavior's value changed. In the second encoding, we represent an event stream with an item in the list for every moment in the time period; for moments at which no event occurs, we insert a special `no-event' symbol. Behaviors are represented by an initial value and a list for which every item represents the behavior's value at a particular timestep. In both encodings, behaviors are discretized rather than truly continuous. Timesteps are logical rather than wall clock time: the precise interval between timesteps need not be specified, but we assume that all inputs are on the same clock; for example, the third value in two input lists are assumed to occur at the same time.

\begin{figure}
    \centering
    \begin{tabular}{ll}
    Event stream & Behavior \\
    \texttt{(10 no-evt 20)} & \texttt{(behavior 0 (10 10 20))} \\
    \texttt{((1 10) (3 20))} & \texttt{(behavior 0 ((1 10) (3 20)))} \\
    \end{tabular}
    \caption{Two possible encodings for event streams and behaviors.}
    \label{fig:encodings}
\end{figure}

Both encodings have algorithmic advantages and disadvantages: for example, the \texttt{mergeE} operator is a sort operation in the explicit timestamp encoding, but a simple linear map operation in the second encoding, while \texttt{delayE} in the second encoding requires a quadratic recursive function to shift events back, but can be calculated in the first encoding by simply adding a constant to all timestamps. However, the second encoding is far more efficient for Rosette. By representing discrete timesteps by the list data structure itself, we halve the number of integers required per inputs, and remove the necessity of any computation to find event ordering or interleaving. This encoding also ensures inputs will always be of the same length, no matter how many events actually occur, whether a filter operation is applied, and so on; this simplifies the implementation of the API.

In our choice of encoding, behaviors are a struct containing an initial value (since behaviors can never have a null value) and a list of the behavior's value at each timestep. Event streams are encoded as a list of event value, or the special symbol indicating no event occurred at that timestep. When event streams are symbolic, they are a list of symbolic unions over a symbolic integer and the special `no event' symbol.

\subsection{Core Flapjax in Rosette}

All that remains is to implement the interpreter for our FRP language. We implement a core of the Flapjax API here, omitting a few of the operators such as \texttt{calmE} and \texttt{blindE}, operators that directly interface with the DOM, and operators that take streams of streams as inputs.\footnote{None of these operators are incompatible with our approach and could be implemented at a later date.} 

The Flapjax behaviors combinators can be modeled as list combinators, just as in the Deepcoder DSL. We extend our semantics to cover event streams as follows. In general, event stream combinators operate only on events. For example, \texttt{andE} emits a true-valued event when both input events are true, false if either or both events are false, and no-event if either or both input streams contain no event at a given timestep. Likewise, combinators that take predicate functions as arguments apply those functions only to events; \texttt{mapE} emits an event that has been applied to the predicate and no-event when no event occurs. For this reason, predicate functions take values as arguments and not events. \texttt{delayE} buffers events and emits them a specified number of timesteps after they occurred; combinators like \texttt{filterRepeatsE} emit only certain events depending on other events that have occurred in the stream. Finally, we can cast between event streams and behaviors using \texttt{changes}, which takes a continuously valued behavior and returns an event stream with events emitted every time the behavior changes value, and \texttt{startsWith}, which takes an initial value and an event stream and fills in every timestep without an event with the value of the most recent event, or the initial value if no event has yet occurred.

In Flapjax, predicate functions are written by the programmer; for the purposes of synthesis, we use the Deepcoder definitions of predicate functions. Program sketches are as in the list DSL encoding above; however, we do not include multiplication or division as it is not required by any of our benchmarks. Our DSL is untyped and using a event operator on a behavior input should throw an error; in practice Rosette is able to easily prune away these kinds of type mismatches.

\begin{figure}
    \centering
    \begin{verbatim}
(define (drag-and-drop-graph 
          mouse-up mouse-down mouse-pos)
  (define r1 mouse-up)
  (define r2 mouse-down)
  (define r3 mouse-pos)
  (define r4 (constantE #f r1))
  (define r5 (constantE #t r2))
  (define r6 (mergeE r4 r5))
  (define r7 (startsWith #f r6))
  (define r8 (snapshotE r3 r7))
  (define r9 (ifE r8 r3 r8))
  (define r10 (filterE (lambda (x) x) r9))
  r10)
  \end{verbatim}
    \caption{The drag and drop program written in straightline register style (Racket).}
    \label{fig:dd-reg}
\end{figure}

We can divide Flapjax operators into two groups: those that are stateless at every timestep and those whose output at each timestep depends on events seen previously. Stateless operators, such as \texttt{constantE} and \texttt{mergeE}, can be independently calculated at each timestep. When given symbolic inputs, they evaluate to formulas linear in size. However, many operators have differing outputs depends on what events have been seen in the past. For example, \texttt{collectE} acts as a fold operator, using a previously calculated value as input to its predicate function, and \texttt{delayE} will emit an event if it has seen an event a specified number of timesteps in the past. To calculate what these operators will emit at a given timestep, we must take into account every event that has already occurred. Thus, when given symbolic inputs, these combinators give formulas that are quadratic in the size of their inputs. 

Since these operators are expensive, we divide the program search space by using templates or metasketches\cite{bornholt2016optimizing} that limit which instruction slots can use them; for example we might first search for a program with no stateful operators, then with a stateful operator in the first instruction slot, and so on. Although we do not do so in this work, executing synthesis using various stateful/stateless operator templates in parallel would be an easy way to improve performance.

\section{The interactive synthesis loop}

With the encoding specified above, we are ready to synthesize programs. All that remains is for our user to provide specifications. Of course, writing specifications is in of itself a significant task; writing a fully constrained specification may be more difficult than writing a program. Our tool allows users to write only a partial specification to begin exploring the space of possible solutions. The full algorithm is specified in Algorithm~\ref{alg:synthloop}.

\begin{algorithm}
  \caption{The interactive synthesis loop algorithm}
  \label{alg:synthloop}
  \begin{algorithmic}[1]
    \Procedure{SynthesisLoop}{specs}
    \State $\textit{P} \gets \text{SynthesizeProgram(}\textit{specs}\text{)}$
    \If {\textit{P} \text{is unsat}} \Return false
    \EndIf
    \State $\textit{P', i} \gets \text{SynthesizeDifferentProgram(}\textit{specs, P}\text{)}$
    \If {\textit{P'} \text{is unsat}} \Return true
    \EndIf
    \If {\textit{i, P(i)} \text{is valid}} \\
    \qquad \Return SynthesisLoop($\{i,P(i)\} \cup specs$)
    \EndIf
    \If {\textit{i, P'(i)} \text{is valid}} \\
    \qquad \Return SynthesisLoop($\{i,P'(i)\} \cup specs$)
    \Else \\
    \qquad \Return SynthesisLoop($\phi \cup specs$)
    \EndIf
    \EndProcedure
  \end{algorithmic}
\end{algorithm}

The simplest form of specification is the input-output pair: the user writes a sample input and the output that the program should produce, as a form of programming by demonstration. When programmers begin to define a function they often have an expected or ``happy path'' execution in mind, and we believe this execution could easily be written down and provided as a partial specification. 

Upon being given this partial specification, the synthesis tool first attempts to synthesize a program that conforms to it. If successful, it then attempts to synthesize a second program that also matches the specification but differs in behavior from the first program on at least one input. As a simple means of encouraging the synthesizer to provide programs that are as different from each other as possible, we simply reorder the list of operators provided to the second sketch, permuting order of clauses in the resulting Rosette formula. After successful synthesis, the user is presented with the two programs, the distinguishing input, and the two differing outputs. The user must then decide how to distinguish between the two programs. The simplest choice would be to add the differing input and the preferred output to the input/output pair specification and run the tool again, continuing the loop until the tool can only find one unique program that conforms to all specifications.

However, writing input/output pairs can be tedious, especially for long executions; it can be easy to make a mistake which will be difficult to later identify. We believe that the user knows other useful facts about the program they are trying to write, and allow them to add these facts to the specification in the form of invariants. There are several types of invariants that can aid program definition. First, the user may provide assumptions about inputs: in the case of the drag and drop example, we may wish to state that a mouse up and mouse down event cannot occur at the same timestep, or that a mouse up event cannot occur until a mouse down event occurs. By providing these assumptions, the user will not be asked to disambiguate programs on invalid inputs, thus eliminating part of the candidate program search space. The user might also provide invariants over outputs, such as stating that the type of the output events is expected to be integers. Finally the user might state invariants over both inputs and outputs, for instance stating that no event should be emitted between a mouse up and a mouse down event. 

\section{Evaluation}

To demonstrate the efficiency of our encoding, we first benchmark program synthesis in the Deepcoder DSL and compare performance with that of two previous implementations. Then, to evaluate our encoding of the FRP paradigm, we have prepared a set of benchmark programs drawn from the web programming and Internet of Things domains. These programs range in size from three to 12 instructions in length and represent realistic tasks. The reference implementations and synthesized programs are both written in Flapjax. First, we demonstrate the size of problem we can handle by synthesizing programs to match a fully constrained specification in the form of reference implementations for each benchmark. Next, we demonstrate the feasibility of the interactive partial specification loop by incrementally adding specifications until we converge on the desired program.

\begin{table*}
\caption{Benchmarks}
{\renewcommand{\arraystretch}{2}
\begin{tabular}{|c|l|}
    \hline
   Benchmark name  &  Description \\
   \hline
   Mousetail  & An element follows the position of the mouse at an offset. \\
   \hline
   Counter buttons & Two buttons increment and decrement a counter. \\
   \hline
   Drag and drop & Click on an element and reposition it with the mouse. \\
   \hline
   Save draft & The text in a field is saved at a given interval if it has changed, or if the user clicks a save button. \\
   \hline
   Thermostat & \begin{minipage}{0.8\textwidth}During the night, turn on the heater if the temperature drops below a given level and turn it off if the temperature rises above a given level.\end{minipage} \\
   \hline
   Sprinklers & \begin{minipage}{0.8\textwidth}Every day at 6pm, turn on the sprinklers for 15 minutes, but only if it has not rained in the last 24 hours. Turn off the sprinklers if motion is detected in the yard. \end{minipage}\\
   \hline
   Kitchen light & \begin{minipage}{0.8\textwidth}Turn on the kitchen light if motion is detected. The color of the light should be white during the day and orange at night. \end{minipage}\\
   \hline
\end{tabular}}
\label{tab:benchmarks}
\end{table*}

\subsection{Deepcoder comparison}

We evaluate synthesis of programs in the Deepcoder DSL and compare our results to two previous efforts. The Deepcoder tool uses machine learning over a large corpus of programs in the list manipulation DSL to learn distributions of combinators, such that for a given a partial program, the next most likely combinator can be guessed. The tool uses this distribution to accelerate a variety of synthesis methods, including a few variations on enumerative search. The Neo synthesis tool uses a CDCL solver with the ability to learn conflict clause lemmas during solving, pruning away portions of the search space. This tool is also benchmarked on the Deepcoder DSL. For comparison, we perform the same benchmarking experiments.

For these experiments, we randomly generate programs with five combinators, taking care to exclude any programs which do not typecheck or programs for which not all combinators affect the return value. We then randomly generate five sets of inputs per program and execute the program to produce their outputs. Finally we use these five input/output pairs as specifications and synthesize a program to match. As an extra step, we then verify whether the synthesized program is semantically equivalent to the target program.

We generated 94 benchmark programs, where half took an integer and a list of integers as input and half took two lists of integers. Input lists were of length 5, and each random integer was chosen from the range -5 to 5. We ran our benchmarks on an EC2 instance of type m4.2xlarge, which has an Intel Xeon E5 processor and 36GB of memory. We set a timeout of one hour; 21 benchmarks timed out and 73 benchmarks ran to completion. Of those that completed, in one case the solver could not find a solution. After we successfully synthesized a solution for each benchmark, we attempted to verify that it was equivalent to the target solution; 32 of the synthesized solutions were semantically equivalent to the original programs, while 40 of them differed. Of the benchmarks that completed, the faster benchmark finished in 87 seconds, while the slowest benchmark took 2605 seconds.

In table~\ref{tab:resultsvsdc}, we find the minimum timeout that would be required to synthesize the benchmarks, broken out by quintile, in order to compare our results with Deepcoder's. Although their results are noticeably faster in the bottom 20\%, we are faster for 40\% of benchmarks and an order of magnitude faster for 60\% of all benchmarks. 

\begin{table}
\caption{The timeouts required to solve 20, 40, 60, and 80\% of all benchmarks, as compared to Deepcoder.}
    \centering
    \begin{tabular}{|l|c|c|c|c|}
    \hline 
       & 20\% & 40\% & 60\% \\
       \hline
       Deepcoder DFS  & 24s & 514s & 2654s \\
       Deepcoder Enumeration  & 9s & 264s & 4640s \\
       Rosette & 113s & 148s & 310s \\
       \hline
    \end{tabular}
    
    \label{tab:resultsvsdc}
\end{table}

\begin{figure*}
    \centering
    \begin{tabular}{cc}
    \includegraphics[width=0.4\textwidth]{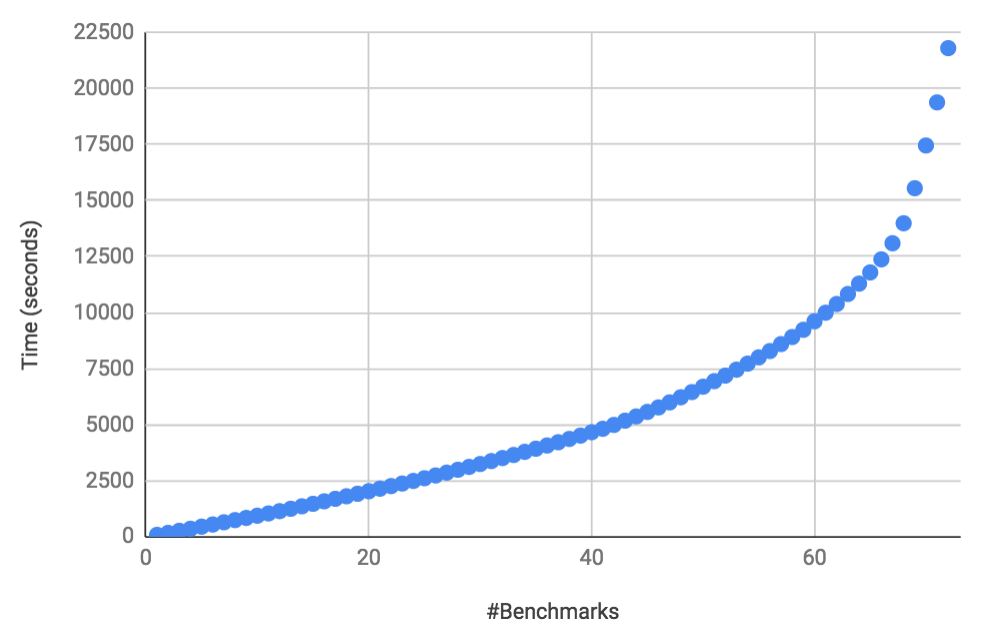} & \includegraphics[width=0.4\textwidth]{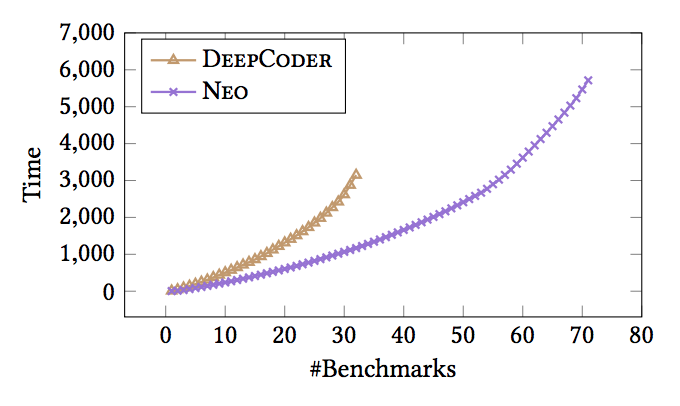}
    \end{tabular}
    \caption{On the left, our results plotted by time; on the right, the results for Neo and Deepcoder (this figure is from Feng et al~\cite{feng2018neo}).}
    \label{fig:resultsvsneo}
\end{figure*}

\subsection{Full specifications}

We have written a reference implementation for each of our benchmarks. We can use these programs as fully constraining specifications and synthesize programs that have the same semantics as the references. This could correspond to the real-world task of synthesizing a shorter or more efficient program that has the same behavior as a longer one (superoptimization), or of synthesizing a program in one language that matches the semantics of a program written in another (transpilation). 

To state the task precisely, for some sketch $S$ which contains some set of holes $H$, we assert there exists some completion of the holes $\vec{h}$ such that for all inputs $i$, when $i$ is applied to the completed sketch, its output is equivalent to the application of $i$ to the target program $P$.

\[ \exists \vec{h} \forall i . S[H := \vec{h}](i) = P(i) \]

\begin{table}
\caption{Results for synthesizing benchmarks against full specifications.}
\begin{tabular}{|c|c|c|c|c|}
\hline
   Benchmark & Insn & Stateful  Insn & AST nodes & Time \\
   \hline
   Mousetail & 3 & 1 & 8 & 1.2s \\
   Increment button & 5 & 2 & 12 & 1.9s \\
   Drag and drop & 7 & 1 & 14 & 13.5s \\
   Save draft & 6 & 2 & 11 & 15.7s \\
   Thermostat & 6 & 0 & 12 & 13.2s \\
   Kitchen light & 13 & 0 & 25 & 497.3s \\
   Sprinklers & 11 & 3 & 16 & 3159.4s \\
   \hline
\end{tabular}

\label{tab:fullspecsynth}
\end{table}

Table~\ref{tab:fullspecsynth} shows the instruction count and number of AST nodes in each benchmark program, and the time it took to synthesize each using the reference implementation as a full specification. Most of the benchmarks are quite fast; note that the kitchen light program, which is twice the size of most of the others, is an order of magnitude slower. The sprinklers benchmark is close to the kitchen light program in size, but uses more stateful operators, resulting in even slower performance. However, these results show that our encoding is capable of synthesizing real programs in a reasonable amount of time.

\begin{figure*}
    \centering
    \begin{tabular}{cc}
    \includegraphics[width=0.5\textwidth]{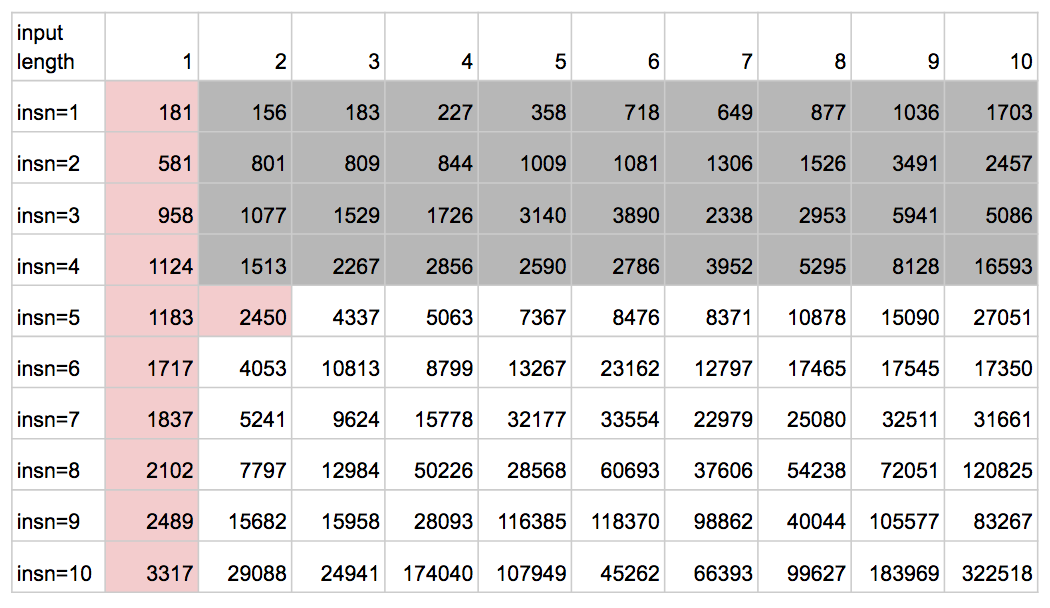} & \includegraphics[width=0.5\textwidth]{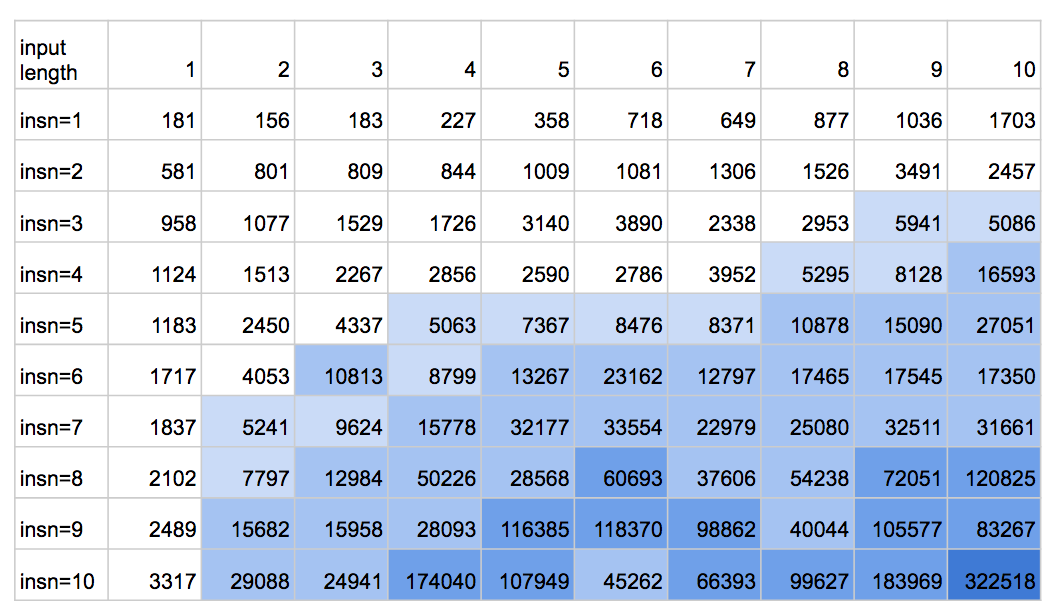}
    \end{tabular}
    \caption{Synthesis times for the drag and drop benchmark using a full specification with different instruction counts and input lengths, showing the best choices for input length and sketch length. On the left, cells shaded red indicate that an incorrect solution was found and cells shaded gray indicate that no solution could be found. On the right, cells are shaded darker blue to indicate longer synthesis times.}
    \label{fig:ddsketchparams}
\end{figure*}

To show the effects of sketch parameters on synthesis performance, we ran the drag and drop benchmark 100 times, using sketches of instruction length 1 through 10 and symbolic input length (representing number of timesteps) 1 through 10. The results are shown in figure~\ref{fig:ddsketchparams}. On the left, fields in gray indicate that no equivalent program could be found with four or fewer instructions. Fields in red indicate where a equivalent program was found on inputs of the specified length, but that the solution is not equivalent to the reference implementation over longer inputs. On the right, shading indicates the length of time it took for synthesis to complete, with darker shading meaning longer synthesis times. We can see that increasing the input length causes the solving time to grow more slowly than increasing the instruction count. For this particular benchmark, a sketch with six instructions over symbolic inputs of length two gives the fastest time. Strategies for choosing sketch parameters is left for future work.

\begin{figure}
\begin{verbatim}
(define (synthesized-function input1 input2 input3)
  (define r1 input1)
  (define r2 input2)
  (define r3 input3)
  (define r4 (constantE #f r1))
  (define r5 (mergeE r4 r2))
  (define r6 (startsWith #f r5))
  (define r7 (snapshotE r3 r6))
  (define r8 (ifE r7 r3 r2))
  r8)
\end{verbatim}
    \caption{The shortest solution that can be synthesized to match the drag and drop benchmark.}
    \label{fig:ddendresult}
\end{figure}

It is worth noting that we are able to find correct programs consisting of five instructions (see figure~\ref{fig:ddendresult}, even though our reference implementation contains seven. The shorter solution takes advantage of the fact that in Racket, all values that are not \texttt{$\#$f} evaluate to true, so there is no need to mask the mouse down events with the constant \texttt{$\#$t}. This is not the case in other languages such as Javascript, so in the future if we target languages other than Racket we will need to faithfully capture these differences in semantics. The other saved instruction comes from a clever choice of the third argument to the final \texttt{ifE} operator; if the program is not in a dragging state and the guard value is false, the mouse down event stream will necessarily be `no-event' and so returns that value, removing the need for a filter.

\subsection{Interactive loop with partial specifications}


\begin{table*}
    \centering
    \caption{A list of interactions leading to the successful synthesis of the drag and drop benchmark.}
    \begin{tabular}{|l|c|c|}
    \hline
      Specification   & Candidate A  & Candidate B \\
      \hline
       Trace: if nothing happens, nothing happens  & Incorrect & Incorrect \\
       Output is of type coordinates & Incorrect & Incorrect \\
       Trace: element is dragged at first timestep, dropped at last & Incorrect & Correct \\
       Mouse up and mouse down cannot occur at same time and must alternate & Incorrect & Incorrect \\
       Trace: alter trace suggested by tool & Incorrect & Incorrect \\
       Trace: alter trace suggested by tool & Incorrect & Correct \\
       Trace: add trace suggested by tool & Correct & - \\
       \hline
       
    \end{tabular}
    \label{tab:ddpartialspecs}
\end{table*}

To demonstrate the use of the interactive tool, we present a series of interactions that results in the successful synthesis of the drag and drop benchmark. Although we wrote the specifications ourselves, we believe them to be reasonable for an end-user to write. The interactions are summarized in table~\ref{tab:ddpartialspecs}; each line lists the new property added to the specification, and whether the two candidate programs the tool produced in response were correct or incorrect. First, we write a trace in which the mouse is never clicked or moved and in response the position of the element never changes. Unsurprisingly, we get back two incorrect programs, one of which returns a series of integers. Our next step is to specify that the output of our function must either be in the coordinate type (a pair of integers representing the x and y position of the element) or `no-event'. Two new incorrect programs are returned, and we add a trace in which the mouse down occurs at the first time step, the mouse up occurs at the last timestep, and the mouse move coordinates appear in the output up until the mouse up occurs. In response we get one correct and one incorrect program, and the tool gives us an invalid input on which the programs have differing behaviors, where mouse up and mouse down events occur at the same timestep. It's not meaningful to define what output the program should have in that instance, so we add a specification stating that mouse up and mouse down events cannot occur at the same time, and that mouse up and mouse down events must alternate. 

Interestingly, we actually get two incorrect candidate programs in response. We take the input suggested by the tool and alter the output to match our desired behavior and try again. This time, we get back one incorrect and one correct program. We choose the suggested input and the correct output and add that to our spec. Finally, the tool reports that there is only one program in its search space that satisfies our specification, and we accept the synthesized program as correct.

For programs in a common domain, such as web programming, we can imagine a library of assumptions that rule out invalid inputs, such as our rule that mouse up and mouse down events cannot occur at the same time. Such properties might be difficult to write for users who are not deeply familiar with browser behavior, but if a domain expert could provide them, they could automatically be included in partial specifications, helping users converge on their desired programs more quickly.

\section{Related Work}

Much has been written about programmers' difficulties with Javascript event handling; see Alimadadi et al. \cite{alimadadi2014understanding} for a discussion of these issues. In the Internet of Things domain, see Brush et al. \cite{brush2011home} for an empirical study of issues users have had with smart devices installed in their homes.

We are not aware of any other work synthesizing functional reactive programs, although as previously mentioned Balog et al. \cite{BalGauBroetal16} and Feng et al. \cite{feng2018neo} use a functional list manipulation language to benchmark their synthesis solvers. However, program synthesis is a well-studied domain. Gulwani 2010 \cite{gulwani2010dimensions} describes many approaches we adopted in this project, including multi-modal specifications to describe user intent and an interactive loop to iteratively refine user specifications; see also Le and Gulwani\cite{le2014flashextract} and other work for synthesis using user-defined input-output examples as specifications.

\section{Future work}

We believe these results show our approach is promising, but much work remains. Future work will include two major directions. First, synthesis performance can be further improved. Users are incredibly sensitive to lag in interactive applications, so our target is to be able to synthesize two candidate programs in less than a second. There are several ways in which we can improve performance of our tool. First, synthesis could easily be parallelized, by issuing synthesis jobs using different metasketches and different parameters. The Flapjax grammar contains a lot of symmetry--for example, \texttt{constantE(3).constantE(2).constantE(1)} is equivalent to \texttt{constantE(1)}, and \texttt{delayE(1).delayE(2)} is equivalent to \texttt{delayE(3)}. Symmetry breaking could greatly reduce the program search space. Further, it's likely that particular groups of instructions often occur together in meaningful programs. If we fused these instructions and added them to the DSL, we could increase the size of the search space in terms of the current DSL without having to increase the number of instructions in our sketches. Finally, as we scale up to more realistic applications, it will become infeasible to synthesize larger programs in a single step. We will need to investigate ways to assist the user in decomposing their programs into smaller components, while making sure that specifications hold over the integrated whole.

The second important direction for future work is usability. At present we synthesize programs by directly writing Racket code; a more usable GUI will need to be created for users. In particular, a graphical tool similar to the Rx marble diagrams\footnote{http://rxmarbles.com/} could greatly assist users in writing accurate input/output traces, and could also be used aid users in understanding the difference between candidate programs. Finally, although the register-style straightline programs are efficient for synthesis, they are difficult for human programmers to read. In order to produce functions that can be incorporated into human-written code, we will need to translate our synthesized programs to a more natural style, or from Racket to another target language such as Javascript.

\section{Conclusion}

In conclusion, we have provided an encoding for the asynchronous datatypes event stream and behaviors, sketches for functional reactive programs, and an interpreter that evaluates those sketches in a DSL based on Flapjax. With these, we define a human-in-the-loop interactive tool that allows users to iteratively define their desired programs by providing partial specifications to a synthesis engine. We show that our encoding achieves good performance in a list manipulation DSL when compared to similar synthesis tools. Using a suite of benchmarks drawn from web programming and Internet of Things programs, we show that we are able to synthesize programs with reasonable performance against full specifications, and provide a sample interaction with the tool to demonstrate how a user might build up specifications to synthesize one of those benchmarks. Finally, we discuss a few directions for future work, including speeding up the performance of the tool and improving the tool's usability.

  

\bibliography{main}


\begin{thebibliography}{11}


\ifx \showCODEN    \undefined \def \showCODEN     #1{\unskip}     \fi
\ifx \showDOI      \undefined \def \showDOI       #1{#1}\fi
\ifx \showISBNx    \undefined \def \showISBNx     #1{\unskip}     \fi
\ifx \showISBNxiii \undefined \def \showISBNxiii  #1{\unskip}     \fi
\ifx \showISSN     \undefined \def \showISSN      #1{\unskip}     \fi
\ifx \showLCCN     \undefined \def \showLCCN      #1{\unskip}     \fi
\ifx \shownote     \undefined \def \shownote      #1{#1}          \fi
\ifx \showarticletitle \undefined \def \showarticletitle #1{#1}   \fi
\ifx \showURL      \undefined \def \showURL       {\relax}        \fi
\providecommand\bibfield[2]{#2}
\providecommand\bibinfo[2]{#2}
\providecommand\natexlab[1]{#1}
\providecommand\showeprint[2][]{arXiv:#2}

\bibitem[\protect\citeauthoryear{Alimadadi, Sequeira, Mesbah, and
  Pattabiraman}{Alimadadi et~al\mbox{.}}{2014}]%
        {alimadadi2014understanding}
\bibfield{author}{\bibinfo{person}{Saba Alimadadi}, \bibinfo{person}{Sheldon
  Sequeira}, \bibinfo{person}{Ali Mesbah}, {and} \bibinfo{person}{Karthik
  Pattabiraman}.} \bibinfo{year}{2014}\natexlab{}.
\newblock \showarticletitle{Understanding JavaScript event-based interactions}.
  In \bibinfo{booktitle}{\emph{Proceedings of the 36th International Conference
  on Software Engineering}}. ACM, \bibinfo{pages}{367--377}.
\newblock


\bibitem[\protect\citeauthoryear{Balog, Gaunt, Brockschmidt, Nowozin, and
  Tarlow}{Balog et~al\mbox{.}}{2017}]%
        {BalGauBroetal16}
\bibfield{author}{\bibinfo{person}{Matej Balog}, \bibinfo{person}{Alexander~L.
  Gaunt}, \bibinfo{person}{Marc Brockschmidt}, \bibinfo{person}{Sebastian
  Nowozin}, {and} \bibinfo{person}{Daniel Tarlow}.}
  \bibinfo{year}{2017}\natexlab{}.
\newblock \showarticletitle{{D}eep{C}oder: {L}earning to Write Programs}.
\newblock \bibinfo{journal}{\emph{5th International Conference on Learning
  Representations (ICLR 2017)}} (\bibinfo{date}{April} \bibinfo{year}{2017}).
\newblock
\urldef\tempurl%
\url{https://arxiv.org/abs/1611.01989}
\showURL{%
\tempurl}


\bibitem[\protect\citeauthoryear{Bornholt, Torlak, Grossman, and Ceze}{Bornholt
  et~al\mbox{.}}{2016}]%
        {bornholt2016optimizing}
\bibfield{author}{\bibinfo{person}{James Bornholt}, \bibinfo{person}{Emina
  Torlak}, \bibinfo{person}{Dan Grossman}, {and} \bibinfo{person}{Luis Ceze}.}
  \bibinfo{year}{2016}\natexlab{}.
\newblock \showarticletitle{Optimizing synthesis with metasketches}. In
  \bibinfo{booktitle}{\emph{ACM SIGPLAN Notices}}, Vol.~\bibinfo{volume}{51}.
  ACM, \bibinfo{pages}{775--788}.
\newblock


\bibitem[\protect\citeauthoryear{Brush, Lee, Mahajan, Agarwal, Saroiu, and
  Dixon}{Brush et~al\mbox{.}}{2011}]%
        {brush2011home}
\bibfield{author}{\bibinfo{person}{AJ Brush}, \bibinfo{person}{Bongshin Lee},
  \bibinfo{person}{Ratul Mahajan}, \bibinfo{person}{Sharad Agarwal},
  \bibinfo{person}{Stefan Saroiu}, {and} \bibinfo{person}{Colin Dixon}.}
  \bibinfo{year}{2011}\natexlab{}.
\newblock \showarticletitle{Home automation in the wild: challenges and
  opportunities}. In \bibinfo{booktitle}{\emph{proceedings of the SIGCHI
  Conference on Human Factors in Computing Systems}}. ACM,
  \bibinfo{pages}{2115--2124}.
\newblock


\bibitem[\protect\citeauthoryear{Czaplicki}{Czaplicki}{2012}]%
        {czaplicki2012elm}
\bibfield{author}{\bibinfo{person}{Evan Czaplicki}.}
  \bibinfo{year}{2012}\natexlab{}.
\newblock \showarticletitle{Elm: Concurrent FRP for functional GUIs}.
\newblock \bibinfo{journal}{\emph{Senior thesis, Harvard University}}
  (\bibinfo{year}{2012}).
\newblock


\bibitem[\protect\citeauthoryear{Fen and Ruben~Martins}{Fen and
  Ruben~Martins}{2018}]%
        {feng2018neo}
\bibfield{author}{\bibinfo{person}{Yu Fen} {and} \bibinfo{person}{Isil~Dillig
  Ruben~Martins, Osbert~Bastani}.} \bibinfo{year}{2018}\natexlab{}.
\newblock \showarticletitle{Program Synthesis using Conflict-Driven Learning}.
  In \bibinfo{booktitle}{\emph{ACM SIGPLAN Notices}}. ACM.
\newblock


\bibitem[\protect\citeauthoryear{Gulwani}{Gulwani}{2010}]%
        {gulwani2010dimensions}
\bibfield{author}{\bibinfo{person}{Sumit Gulwani}.}
  \bibinfo{year}{2010}\natexlab{}.
\newblock \showarticletitle{Dimensions in program synthesis}. In
  \bibinfo{booktitle}{\emph{Proceedings of the 12th international ACM SIGPLAN
  symposium on Principles and practice of declarative programming}}. ACM,
  \bibinfo{pages}{13--24}.
\newblock


\bibitem[\protect\citeauthoryear{Le and Gulwani}{Le and Gulwani}{2014}]%
        {le2014flashextract}
\bibfield{author}{\bibinfo{person}{Vu Le} {and} \bibinfo{person}{Sumit
  Gulwani}.} \bibinfo{year}{2014}\natexlab{}.
\newblock \showarticletitle{FlashExtract: a framework for data extraction by
  examples}. In \bibinfo{booktitle}{\emph{ACM SIGPLAN Notices}},
  Vol.~\bibinfo{volume}{49}. ACM, \bibinfo{pages}{542--553}.
\newblock


\bibitem[\protect\citeauthoryear{Meyerovich, Guha, Baskin, Cooper, Greenberg,
  Bromfield, and Krishnamurthi}{Meyerovich et~al\mbox{.}}{2009}]%
        {meyerovich2009flapjax}
\bibfield{author}{\bibinfo{person}{Leo~A Meyerovich}, \bibinfo{person}{Arjun
  Guha}, \bibinfo{person}{Jacob Baskin}, \bibinfo{person}{Gregory~H Cooper},
  \bibinfo{person}{Michael Greenberg}, \bibinfo{person}{Aleks Bromfield}, {and}
  \bibinfo{person}{Shriram Krishnamurthi}.} \bibinfo{year}{2009}\natexlab{}.
\newblock \showarticletitle{Flapjax: a programming language for Ajax
  applications}. In \bibinfo{booktitle}{\emph{ACM SIGPLAN Notices}},
  Vol.~\bibinfo{volume}{44}. ACM, \bibinfo{pages}{1--20}.
\newblock


\bibitem[\protect\citeauthoryear{Newcomb, Chandra, Jeannin, Schlesinger, and
  Sridharan}{Newcomb et~al\mbox{.}}{2017}]%
        {newcomb2017iota}
\bibfield{author}{\bibinfo{person}{Julie~L Newcomb}, \bibinfo{person}{Satish
  Chandra}, \bibinfo{person}{Jean-Baptiste Jeannin}, \bibinfo{person}{Cole
  Schlesinger}, {and} \bibinfo{person}{Manu Sridharan}.}
  \bibinfo{year}{2017}\natexlab{}.
\newblock \showarticletitle{IOTA: a calculus for internet of things
  automation}. In \bibinfo{booktitle}{\emph{Proceedings of the 2017 ACM SIGPLAN
  International Symposium on New Ideas, New Paradigms, and Reflections on
  Programming and Software}}. ACM, \bibinfo{pages}{119--133}.
\newblock


\bibitem[\protect\citeauthoryear{Torlak and Bodik}{Torlak and Bodik}{2014}]%
        {torlak2014lightweight}
\bibfield{author}{\bibinfo{person}{Emina Torlak} {and}
  \bibinfo{person}{Rastislav Bodik}.} \bibinfo{year}{2014}\natexlab{}.
\newblock \showarticletitle{A lightweight symbolic virtual machine for
  solver-aided host languages}. In \bibinfo{booktitle}{\emph{ACM SIGPLAN
  Notices}}, Vol.~\bibinfo{volume}{49}. ACM, \bibinfo{pages}{530--541}.
\newblock


\end{thebibliography}

\appendix
\section{Appendix}\label{sec.dcdslgrmr}

\subsection{List manipulation grammar}

\begin{grammar}
<list-expr> ::= <list> 
\alt `take' <number-expr> <list-expr>
\alt `drop' <number-expr> <list-expr>
\alt `reverse' <list-expr>
\alt `sort' <list-expr>
\alt `map' <number-predicate> <list-expr>
\alt `filter' <bool-predicate> <list-expr>
\alt `count' <bool-predicate> <list-expr>
\alt `zipwith' <number2-predicate> <list-expr> <list-expr>
\alt `scanl1' <number2-predicate> <list-expr>

<number-expr> ::= <number>
\alt `access' <number-expr> <list-expr>
\alt `minimum' <list-expr>
\alt `maximum' <list-expr>
\alt `sum' <list-expr>

<number-predicate> ::= (+ 1) | (- 1) | (* 2)
\alt (/ 2) | (* (- 1)) | (** 2) | (* 3)
\alt (/ 3) | (* 4) | (/ 4)

<bool-predicate> ::= `positive?' | `negative?'
\alt `odd?' | `even?'

<number2-predicate> ::= + | - | * | `min' | `max'
\end{grammar}

\subsection{Core Flapjax grammar}

\begin{grammar}
<event-stream> ::= `identityE'
\alt `onceE'
\alt `zeroE'
\alt `mapE' <int-predicate> <event-stream>
\alt `mergeE' <event-stream> <event-stream>
\alt `filterE' <bool-predicate> <event-stream>
\alt `ifE' <bool-event-stream> <event-stream> <event-stream>
\alt `constantE' <constant> <event-stream>
\alt `collectE' <constant> <int2-predicate> <event-stream>
\alt `filterRepeatsE' <event-stream>
\alt `snapshotE' <event-stream> <behavior>
\alt `delayE' <constant> <event-stream>
\alt `timerE' <constant> <event-stream>
\alt `changes' <behavior>
\alt <bool-event-stream>

<bool-event-stream> ::= `andE' <bool-event-stream> <bool-event-stream>
\alt `orE' <bool-event-stream> <bool-event-stream>
\alt `notE' <bool-event-stream>

<behavior> ::= `startsWith' <constant> <event-stream>
\alt `constantB' <constant>
\alt `delayB' <constant> <behavior>
\alt `liftB' <int-predicate> <behavior>
\alt `ifB' <bool-behavior> <behavior> <behavior>
\alt `timerB' <constant> <behavior>
\alt <bool-behavior>

<bool-behavior> ::= `andB' <bool-behavior> <bool-behavior>
\alt `orB' <bool-behavior> <bool-behavior>
\alt `notB' <bool-behavior> <bool-behavior>

<int-predicate> ::= ($\lambda$ (i) (+ i <constant>))
\alt ($\lambda$ (i) (- i <constant>))
\alt ($\lambda$ (i) (- <constant> i))
\alt ($\lambda$ (i) (* i <constant>))

<bool-predicate> ::= ($\lambda$ (i) ($\leq$ i <constant>))
\alt ($\lambda$ (i) ($\geq$ i <constant>))
\alt ($\lambda$ (i) (< i <constant>))
\alt ($\lambda$ (i) (> i <constant>))
\alt ($\lambda$ (i) (or ($\geq$ i <constant>) ($\leq$ i <constant>)))
\alt ($\lambda$ (i) (and ($\geq$ i <constant>) ($\leq$ i <constant>)))

<int2-predicate> ::= + | - | `min' | `max'
\end{grammar}

\end{document}